\documentclass[sigconf]{acmart}
\makeatletter
\def\@ACM@checkaffil{
    \if@ACM@instpresent\else
    \ClassWarningNoLine{\@classname}{No institution present for an affiliation}%
    \fi
    \if@ACM@citypresent\else
    \ClassWarningNoLine{\@classname}{No city present for an affiliation}%
    \fi
    \if@ACM@countrypresent\else
        \ClassWarningNoLine{\@classname}{No country present for an affiliation}%
    \fi
}
\makeatother
\usepackage{graphicx}
\usepackage{caption}
\usepackage{subcaption}
\usepackage{hyphenat}
\usepackage{tikz}

\usepackage{multicol}
\usepackage{setspace}
\usepackage{lipsum}

\AtBeginDocument{%
  \providecommand\BibTeX{{%
    \normalfont B\kern-0.5em{\scshape i\kern-0.25em b}\kern-0.8em\TeX}}}

\copyrightyear{2023}
\acmYear{2023}
\setcopyright{acmlicensed}\acmConference[L@S '23]{Proceedings of the Tenth ACM Conference on Learning @ Scale}{July 20--22, 2023}{Copenhagen, Denmark}
\acmBooktitle{Proceedings of the Tenth ACM Conference on Learning @ Scale (L@S '23), July 20--22, 2023, Copenhagen, Denmark}
\acmPrice{15.00}
\acmDOI{10.1145/3573051.3596185}
\acmISBN{979-8-4007-0025-5/23/07}

\begin{document}

\title[Envisioning an Inclusive Metaverse]{Envisioning an Inclusive Metaverse: Student Perspectives on Accessible and Empowering Metaverse-Enabled Learning}
\titlenote{This paper has been accepted for presentation at the L@S 2023 conference. The version provided here is the pre-print manuscript.}

\author{Reza Hadi Mogavi}
\authornote{Corresponding Authors}
\affiliation{%
  \institution{HKUST and University of Waterloo}
}
\email{rhadimogavi@acm.org}

\author{Jennifer Hoffman}
\authornotemark[1]
\affiliation{%
  \institution{Accessible Meta Group}
}
\email{jhoffmang@accessiblemeta.org}

\author{Chao Deng}
\affiliation{%
  \institution{Accessible Meta Group}
}
\email{cdeng@accessiblemeta.org}

\author{Yiwei Du}
\affiliation{%
  \institution{Dalian University of Technology}
}
\email{duyiwei@mail.dlut.edu.cn}

\author{Ehsan-Ul Haq}
\affiliation{%
  \institution{Hong Kong University of Science and Technology (Guangzhou)}
}
\email{euhaq@connect.ust.hk}

\author{Pan Hui}
\authornotemark[1]
\affiliation{%
  \institution{Hong Kong University of Science and Technology (Guangzhou)}
}
\email{panhui@ust.hk}

\begin{abstract}
The emergence of the metaverse is being widely viewed as a revolutionary technology owing to a myriad of factors, particularly the potential to \textit{increase the accessibility of learning for students with disabilities}. However, not much is yet known about the views and expectations of disabled students in this regard. The fact that the metaverse is still in its nascent stage exemplifies the need for such timely discourse. To bridge this important gap, we conducted a series of semi-structured interviews with 56 university students with disabilities in the United States and Hong Kong to understand their views and expectations concerning the future of metaverse-driven education. We have distilled student expectations into five thematic categories, referred to as the \textit{REEPS} framework: \textit{Recognition}, \textit{Empowerment}, \textit{Engagement}, \textit{Privacy}, and \textit{Safety}. Additionally, we have summarized the main design considerations in eight concise points. This paper is aimed at helping technology developers and policymakers plan ahead of time and improving the experiences of students with disabilities.
\end{abstract}




\maketitle

\section{Introduction}
The realization of the universal \textit{metaverse} gets closer with each passing day \cite{10.1145/3491101.3516399, 10.1145/3491101.3519779, 10.1145/3491102.3502008, 10.1145/3491102.3502082, 10.1145/3474085.3479238}. Although scholars have defined the term \textit{metaverse} in myriad ways (see \cite{lee2021all, Tlili2022}), they are yet to settle on a single definition. For this paper, we adhere to the simplest and broadest definition of the term, describing it as the next generation of the Internet in which people (represented by avatars) would be able to live, work, study, play, and interact with each other and software applications in virtual worlds \cite{10.1145/3474085.3479238, lee2021all}.

Although virtual and augmented reality (VR/AR) are often seen as the foundation of metaverse technology \cite{lee2021all, hoffert2022metaverse}, it is important to understand that the concept of metaverse extends beyond these technologies. In fact, the metaverse aims to create a novel and inclusive sociotechnical system with its unique social and engineering properties, such as \textit{persistence}, \textit{synchronicity}, and unprecedented \textit{interoperability} \cite{10.1145/3532525.3532534, lee2021all, hoffert2022metaverse}.

Despite ongoing debates about the definition of the term metaverse \cite{lee2021all, Tlili2022}, numerous large companies have already established their own branches within the metaverse, including \textit{Decentraland} \cite{DecentralandSite}, \textit{Spatial} \cite{SpatialSite}, and \textit{Mozilla Hubs} \cite{MozillaHubsSite}.\footnote{These metaverse platforms often promote themselves as gateways to the ultimate universal metaverse, which is expected to be established in the future \cite{lee2021all}.} In 2021, Meta (formerly Facebook) invested over \$10 billion in developing metaverse technologies, and plans to invest even more in the coming years \cite{10.1145/3546607.3546611}. \textit{Bloomberg Intelligence} estimates that the metaverse market will be worth approximately \$800 billion by 2024 \cite{BloombergStat}. Against this backdrop, a wide range of industries and businesses, including \textit{music}, \textit{fashion}, \textit{food}, \textit{games}, and \textit{social media}, are seeking to capitalize on the untapped potential of this new technology \cite{10.1145/3491101.3516399, 10.1145/3491101.3519779, 10.1145/3491102.3502008, 10.1145/3491102.3502082, 10.1145/3474085.3479238, lee2021all}.

\begin{figure*}[t!]
\centering
    \begin{minipage}[b]{0.48\textwidth}
        \centering
        \includegraphics[width=\textwidth]{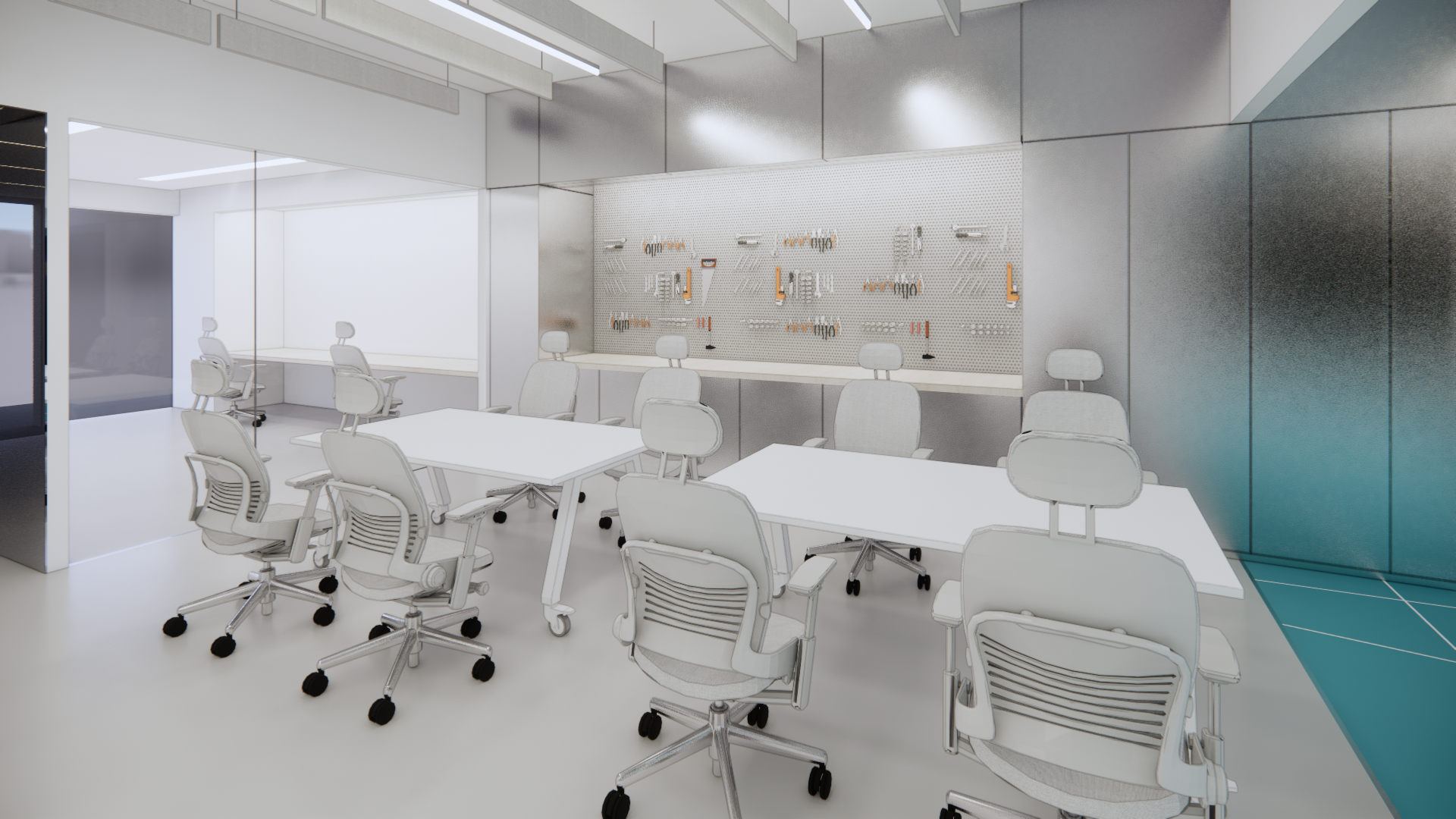}
        \subcaption{A futuristic, metaverse-based classroom in operation at the university (metaversity) associated with the last author}~\label{fig:realclass}
    \end{minipage}\hfill
    \begin{minipage}[b]{0.48\textwidth}
        \centering
        \includegraphics[width=\textwidth]{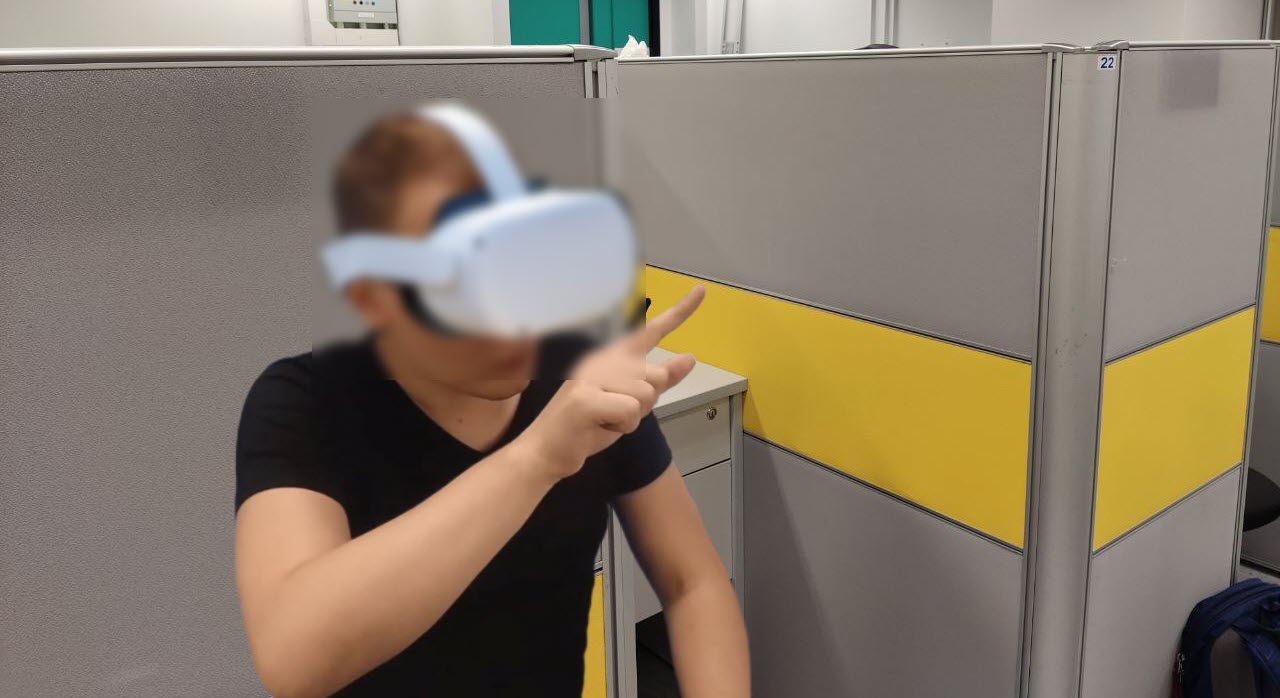}
        \subcaption{A disabled student who is using their VR headset to attend a metaverse-based classroom}~\label{fig:realstudent}
     \end{minipage}
\caption{Figure (a) shows an illustration of a futuristic metaverse-based classroom set up at the last author's university. Figure (b) shows an anonymized photo of a student (with autism) who is using their VR headset (i.e., Oculus Quest 2) to attend a virtual lab session. Permission for using this photo has already been obtained from the student.}~\label{fig:realmetaverseclassroom}
\end{figure*}

\textbf{Education} is widely considered one of the most promising areas for the application of metaverse technology \cite{Tlili2022, 10.1145/3514262.3514345}. In recent years, several pioneering institutions of higher education have created and launched their own unique \textit{digital twin} branches on metaverse platforms like Spatial and Mozilla Hubs \cite{metaversitiesSite, lee2021all, Tlili2022}. Digital twins are virtual replicas of real-world places and entities that enable seamless interoperability and data transmission between the physical and virtual worlds \cite{Wang2022}. These metaverse-based replicas are also commonly referred to as ``metaversities'' or meta universities. Figure \ref{fig:realclass} displays an example of a futuristic metaverse-based classroom currently in use and under further development at the last author's affiliated university (or metaversity). These classrooms aim to enhance learning experiences for all students through hands-on and immersive methods. Figure \ref{fig:realstudent} shows an anonymized photo of an autistic student using a VR headset to attend a metaverse-based lab session.

As the metaverse's popularity continues to expand in the realm of education and learning at scale (L@S), the topics of \textit{inclusion} and \textit{accessibility} have also become more prominent than ever \cite{Tlili2022, 10.1145/3491102.3517542, 10.1145/3474085.3479238}. It is imperative to make sure that students with disabilities have a voice in discussions about the future of higher education in the metaverse \cite{Tlili2022, 10.1145/3532525.3532534}: resonating with the tagline``\textit{nothing about us without us!}'' The fact that metaverse-based education is still in its early stages exemplifies the need for such timely discourse, given that it can possibly lead to more effective and cost-efficient accessibility solutions than when technology is fully established and is far too pervasive to be altered \cite{world2015global}. That said, there is still a paucity of scholarly research on this topic, and we know little about the views and expectations of students with disabilities regarding education in the metaverse (see the latest literature review conducted by Tlili et al. \cite{Tlili2022}). In this study, we want to take the first steps toward filling this significant gap in the literature. 

To this end, we conducted one-on-one interviews with 56 university students with disabilities in the United States as well as Hong Kong to learn about their views and expectations about the future of metaverse-driven education. According to our findings, most students were \textbf{optimistic and enthusiastic} about the metaverse's new educational possibilities, and they believed that it could help close some of the current educational gaps between disabled and non-disabled students. For example, some students noted that avatars in the metaverse could help them find teammates for doing group projects more easily than in offline classrooms. In most traditional offline classrooms, they think that they are often judged hastily and unfairly based only on how they appear (rather than their capabilities and merits). Having said that, the students also expressed \textbf{grave concerns about being forgotten and denied their educational rights again}, similar to what they had previously experienced when taking online classes during the COVID-19 pandemic. Most students also emphasized the importance of dispelling a common misconception among educators and educational technology developers. This misconception is that digitization will automatically, or ``\textit{magically},'' make education more accessible. They insisted that a more accessible learning environment cannot be achieved unless this misconception is addressed. Therefore, this paper aims to underscore that merely relying on digitization is insufficient for achieving greater accessibility in education.

Utilizing a careful thematic analysis, we synthesized students' expectations of an ideal metaverse classroom into five key thematic categories, which we henceforth refer to as \textit{REEPS}, standing for \textit{Recognition}, \textit{Empowerment}, \textit{Engagement}, \textit{Privacy}, and \textit{Safety}. In all our interviews, we repeatedly heard about the need for the metaverse to recognize and empower students with various types of disabilities (i.e., motor, visual, cognitive, and hearing impairments). The students also stressed many times that transitioning to the realm of the metaverse alone is insufficient. They said that educators and developers should also think about ways to meaningfully engage them with the learning materials, their peers, and the metaverse itself. Finally, some students expressed concerns about privacy and the safety of the metaverse for students with disabilities, asking for further assurances.

\textbf{Contributions.} To the best of our knowledge, this paper is the first qualitative research that explores the views and expectations of students with disabilities about the future of higher education in the metaverse. It contributes to the Human-Computer Interaction (HCI) and L@S communities in three ways: (1) It expounds upon REEPS, which synthesizes the perspectives of a relatively large number of students with disabilities on the core values they appreciate the most in the future development of metaverse-driven education. (2) We present an insightful set of design considerations so that technology developers as well as policymakers can plan ahead of time and improve the experiences of students with disabilities. (3) Finally, our paper helps create a timely conversation space in the HCI and L@S communities on the critical topics of inclusion and accessibility in metaverse-driven education given that the technology is still in its infancy.


\section{Method}\label{sec:method}
This section provides a brief overview of our research methodology and outlines the key details of our study.
\subsection{Participants} 
For this study, participants were recruited from 21 disability centers situated in the United States, with a particular focus on the state of Michigan. Additionally, four centers located in Hong Kong were also included in the recruitment process. The means by which this recruitment took place included bulletin board advertisements, newsletters (sent through email), as well as social media channels on Facebook, Twitter, and Reddit. We also leveraged snowball sampling to maximize the reach of our advertisements \cite{blandford2016qualitative}. The study required the participants to fill out an initial-screening questionnaire prepared using Qualtrics \cite{QualtricsSite}. The questionnaire included inquiries regarding the participants' demographic details, the type of disability they have, and their level of familiarity with the metaverse and its associated applications on a 5-point Likert scale.

The inclusion criteria for selecting participants required them to: (1) be aged at least 18 years old;  (2) be studying at an accredited higher education institution; (3) have at least one type of disability; and (4) be acquainted with the concept of the metaverse prior to the commencement of this study. Specifically, they had to be familiar with at least one metaverse platform, for instance, Decentraland.\footnote{The last criterion was included to ensure the selection of informant participants, regardless of whether or not they had practical experience with it.}

This study ensured compliance with the principle of maximal variation \cite{10.1145/3411764.3445737} by recruiting participants from a variety of demographic and disability backgrounds. This approach helped us figure out if the voices of disabled students shared any common perspectives collectively. Further details can be found in Appendix \ref{sup1}.
   
\subsection{Semi-Structured Interviews}
After obtaining ethical approval from the Institutional Review Board (IRB) in our university, we gathered signed consent forms from the participants and/or their legal caregivers\footnote{Permission from the legal caregivers was only obtained if the participant was unable to provide it themselves, or if the law required it.}  through email. In light of the high number of COVID-19 cases (Omicron variant) during the period of this study and to ensure the safety of our participants, we decided to conduct all interviews online using their preferred video conferencing applications, such as Webex, Google Duo, Zoom, or Skype.

The interviews were conducted by the first three authors, who contributed equally to the interviews' planning and execution. All three interviewers have received formal academic training in working with human subjects and human data protection. They also have experience working closely with students with disabilities. Further details can be found in Appendix \ref{sup2}.

\subsection{Data Analysis}
The first two authors of this study (henceforth referred to as coders) collaboratively analyzed the interview data for two months by utilizing an inductive coding process \cite{gavin2008understanding}. This coding style was selected because it is more congruent with the exploratory nature of our study (also see \cite{10.1145/3449168, mogavi2022promotional}). In this study, the coders have employed a paid version of the qualitative analysis application known as ATLAS.ti\footnote{https://atlasti.com/} to manage, organize, and exchange their findings in a more systematic and recorded manner. 

Following multiple rounds of discussion and collaboration, a consensus was reached for the final codebook. Both coders ultimately agreed on five recurring main themes: Recognition, Empowerment, Engagement, Privacy, and Safety. Additional details can be found in Appendix \ref{sup3}.

\section{Findings and Discussion}
This section presents the primary findings of our work and discussions across three subsections: student impressions, the REEPS framework, and design considerations.

\subsection{Student Impressions}
Based on our research, it appears that the majority of students (n = 40, 71\%) are enthusiastic and optimistic about the educational opportunities that are going to be enabled by the metaverse. P4, a disabled student with limited physical mobility, expressed excitement about the possibilities of the metaverse: ``\textit{I think the metaverse is a good opportunity, and I can't wait to explore all of the new possibilities it will bring to us. I'm especially excited about the virtual classroom experience, which I think will be more accommodating than the traditional classrooms. ... I think it will be fantastic to have the ability to move my body in ways that I've always wanted to, such as running, jumping, and participating in various immersive virtual sports.}'' 

However, some students (n = 12, 21\%) worry about being forgotten and denied their educational rights, replacing real teachers with virtual ones, unequal access to resources, isolation, privacy, and safety concerns. One student, P8, voiced their concern, saying, ``\textit{I understand the potential benefits of the metaverse, but I worry that it'll be another way for schools to cut corners and save money. What if they decide to replace real teachers with virtual ones? What if they stop providing special services for students with disabilities, thinking that the metaverse can solve everything? And some families might not be able to afford the customized equipment and technology needed to participate in the metaverse, which could lead to inequities and exclusion. I don't want to go back to being invisible behind digital walls.}''

Four students expressed indifference towards the emergence of the metaverse for two main reasons: two felt that it was a passing fad, while the other two preferred to continue with either face-to-face or online classes. Metaverse is probably not going to change everything anyway.
\subsection{REEPS Framework: Moving Toward an Inclusive Metaverse}
\begin{figure}[t!]
     \centering
    \begin{minipage}[b]{0.45\textwidth}
         \centering
         \includegraphics[width=0.5\textwidth]{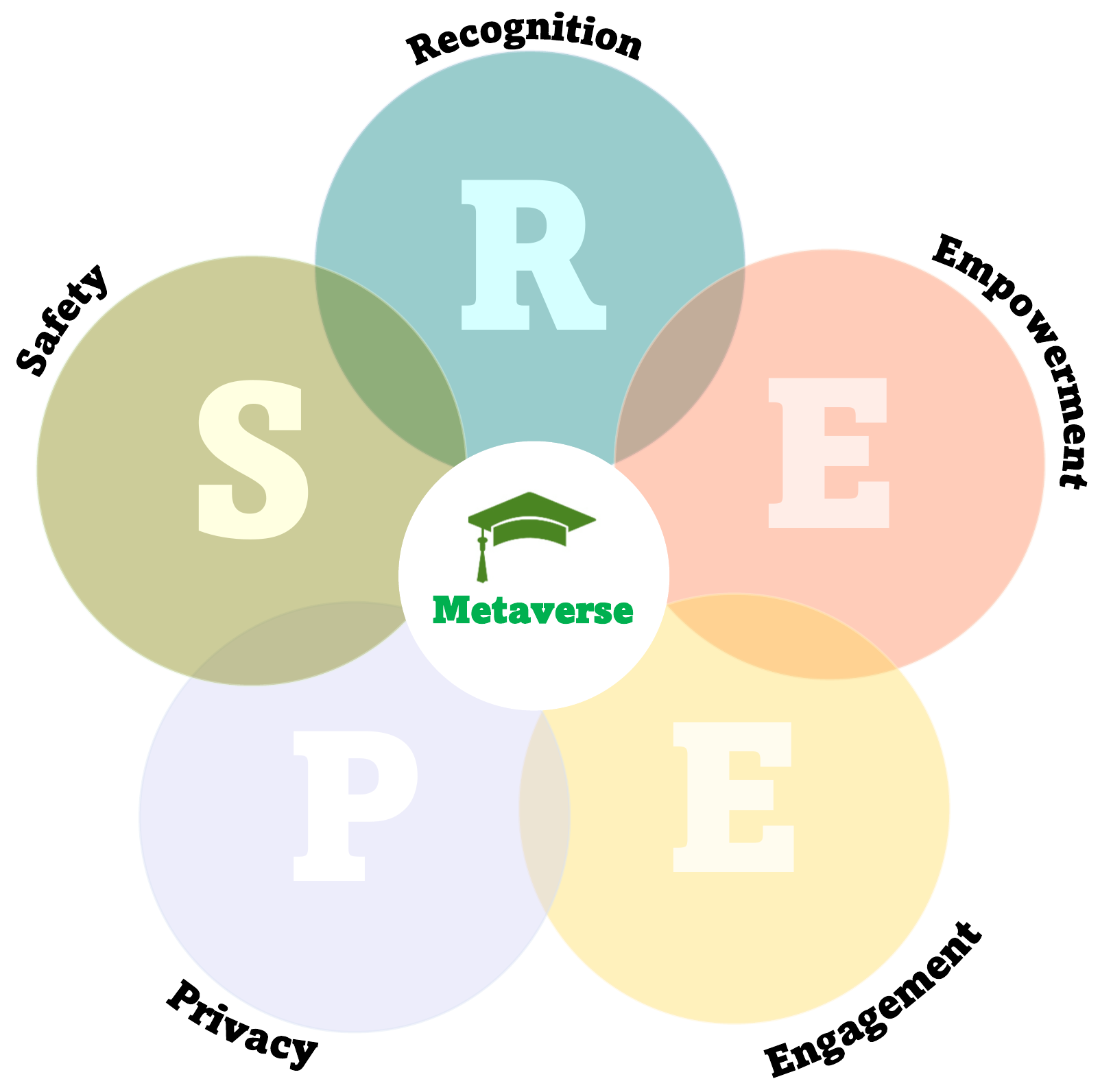}
     \end{minipage}
        \caption{The REEPS framework for Inclusive Metaverse-Enabled Education: Recognition, Empowerment, Engagement, Privacy, and Safety,}~\label{fig:reepsframework}
    \end{figure}

To create an inclusive metaverse, it's crucial to understand the needs and expectations of students with disabilities. The REEPS framework for Inclusive Metaverse-Enabled Education (Figure \ref{fig:reepsframework}) provides a comprehensive set of principles for meeting these needs and expectations.
\begin{itemize}
    \item \textbf{Recognition:} Students request that tech companies recognize disability inclusion and accommodation as a high priority, not an afterthought. They note that having a disability does not imply inferiority or reduced capabilities. Fostering a proper understanding and accommodation of disabilities is a crucial initial step toward creating an inclusive metaverse.
    \item \textbf{Empowerment:} Based on our findings, numerous students advocate for opportunities in the metaverse that enable self-directed learning and customization of experiences. They aspire to be creators and equal contributors rather than passive users. Providing universal access and tools to promote independence is crucial for empowering students with disabilities.
    \item \textbf{Engagement:} The aspect of engagement is a critical consideration in the development and implementation of the metaverse as a learning platform. Our research has shown that an immersive, active, and socially connected metaverse is essential for promoting active student participation and facilitating interaction with both instructors and peers in a barrier-free manner \cite{10.1145/3430895.3460126}. To foster engagement, it is crucial to explore virtual activities tailored to the diverse needs and interests of students \cite{mogavi2023your, 10.1145/3449086}. Additionally, transcending the limitations of the physical world is imperative to unlock the full potential of the metaverse as a transformative educational tool.
    \item \textbf{Privacy:} Many students insist on controlling their data and identity in the metaverse. Personal information should be disclosed at users' discretion to ensure comfort and security.
    \item \textbf{Safety:} The metaverse should place a high value on safety across various levels, such as the physical environment, physical equipment, virtual environment, and virtual applications. Our research shows that many students require a metaverse-based education where harassment and abuse based on disabilities are not tolerated. Clear rules, reliable reporting systems, and an inclusive culture are essential to achieve this.
\end{itemize}

\subsection{Practical Design Considerations: Eight Conclusive Points}
When designing for disabled students, it is important to consider several practical design considerations. Firstly, it is crucial to recognize and challenge misconceptions about disability while also educating oneself about various disabilities. Secondly, non-apparent disabilities such as autism should also be taken into account. Thirdly, gathering feedback from a diverse range of people is essential. Fourthly, accessibility solutions should be both affordable and designed to maintain security and privacy. Fifthly, both physical and virtual worlds should support assistive technologies. Sixthly, universal design principles should be applied to allow for customization and future plugins. Seventhly, the collaboration between students, teachers, peers, and caregivers should be facilitated. Finally, accessibility solutions should be continuously evaluated and improved through an iterative process to ensure that the needs of disabled students are being met.

\section{Final Remarks}

According to our findings, inclusive and accessible metaverse-driven education is not possible unless technology designers and developers recognize and understand the need to establish REEPS qualities as a \textit{social necessity} rather than an ``\textit{afterthought}'' or ``\textit{a kind act or nice gesture}.'' This is an important point because, according to the literature, most designers and developers today tend to routinely overlook the needs of people with disabilities when developing their products and focus solely on non-disabled individuals as the only \textit{de facto} target audience \cite{10.1145/2982142.2982158, 10.1145/3234695.3236346}. Most students mentioned that designers and developers might have the wrong assumption that digitization alone will automatically make education more accessible. 

As metaverse and extended reality (XR) technologies emerge, recognizing disabilities, especially non-apparent ones (e.g., autism), becomes more challenging. Hence, it is important to fully acknowledge diversity and expression.

Our findings highlight an ``\textit{accessibility impasse}'' as a significant challenge, where technology developers and higher education institutions are unclear about who is responsible for providing certain accessibility features, \textbf{leading to inaction}. Developers may think it is the institutions' or students' responsibility, while institutions may believe it falls on developers or students themselves.

We hope the findings of this paper underscore the necessity for tech designers to prioritize accessibility and cater to the diverse needs of students with disabilities within metaverse-driven education.

\begin{acks}
The authors gratefully acknowledge the support, in part, from the MetaHKUST project at the Hong Kong University of Science and Technology (Guangzhou) for this research.
\end{acks}

\bibliographystyle{ACM-Reference-Format}
\bibliography{sample-base}

\appendix
\section{Supplementary Details on Research Participants}\label{sup1}
We kept recruiting new participants until we did not hear any new points of view from them; at that point, we determined that we had collected enough data (data saturation). We recruited 56 university students aged 19-37 years (median = 25 years), of whom 32 were identified as female, with the number of male participants being 24. They came from a variety of ethnic groups: White (n = 11), Black or African American (n = 10), Hispanic and Latino (n = 7), Asian (n = 9), Native American (n = 9), mixed race (n = 6), and other races (n = 4). Our participants had different types of disabilities that fall into the categories of motor (n = 20), visual (n = 18), cognitive (n = 15), and hearing impairments (n = 10). Notably, the distribution of disabilities does not add up to 56 because some participants had multiple impairments at the same time: motor and visual (n = 3), cognitive and visual (n = 2), motor and hearing (n = 1), and hearing and visual (n = 1).

Our study reports on the perspectives of both local and international students from the United States and Hong Kong. In this study, there were a total of 38 students from the United States (local (native) = 10, international = 28) and 18 from Hong Kong (local = 7, international = 11). 44 participants reported being born with their impairments, while the remaining 12 reported having acquired their conditions at a later stage in life. The participants were inclusive of 46 undergraduate students (humanities and social sciences = 32, science and engineering = 9, other streams = 5) and 10 postgraduate students (humanities and social sciences = 7, science and engineering = 2, other streams = 1). Nevertheless, only 21 participants had prior experience with immersive technologies and metaverse platforms in a real classroom setting. The rest of the students either lacked practical experience (n = 25) or utilized other means to derive immersive metaverse experiences, e.g., VR/AR-based therapy sessions, meta-based exercises (for fitness), rapid product testing (during some accessibility experiments), and a couple of immersive metaverse-based games. 

\begin{figure}[t!]
     \centering
    \begin{minipage}[b]{0.45\textwidth}
         \centering
         \includegraphics[width=\textwidth]{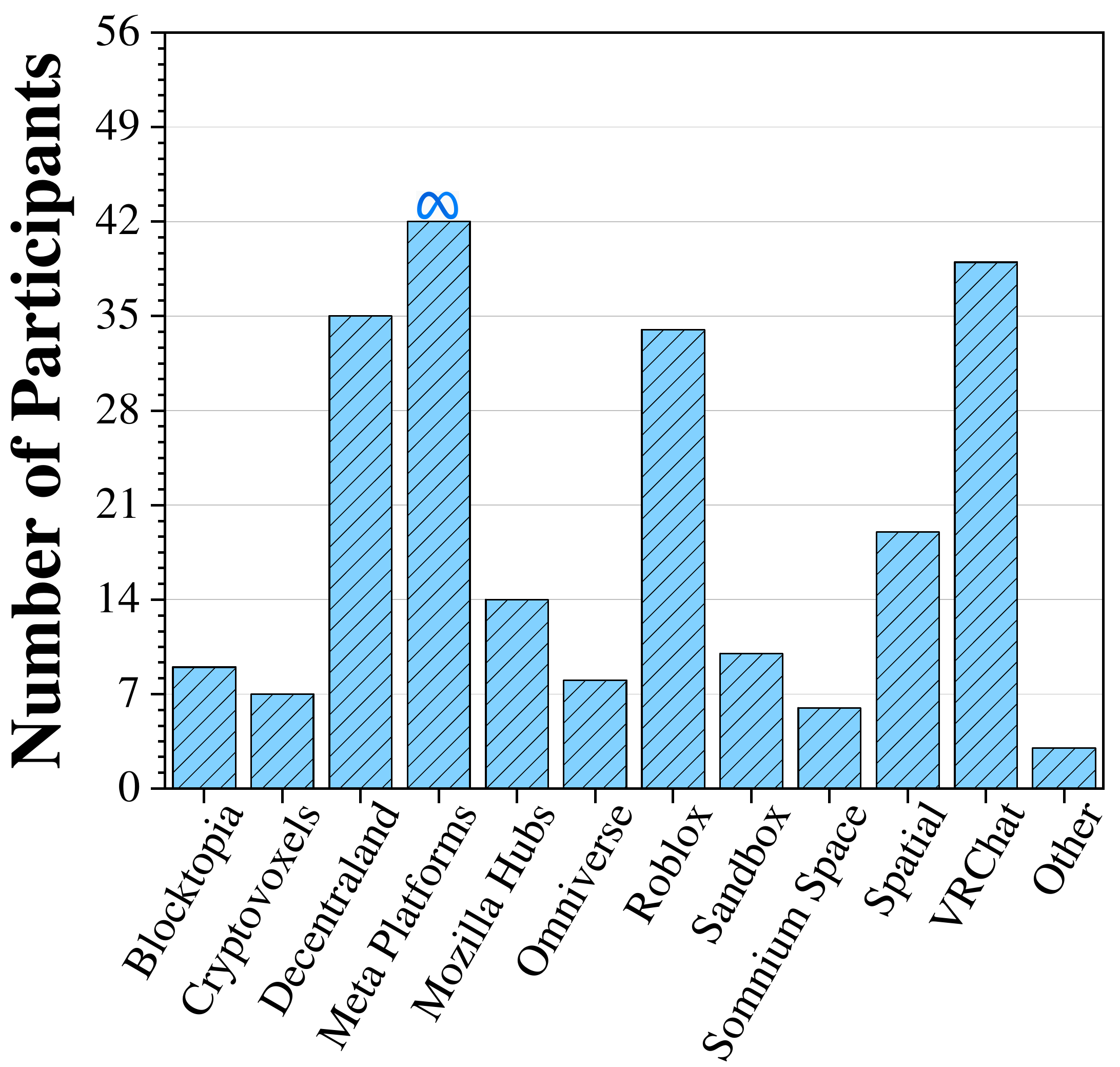}
     \end{minipage}
        \caption{Number of participants familiar with each of the mainstream metaverse platforms in 2022. The numbers are not mutually exclusive.}~\label{fig: dist1}
    \end{figure}

According to the findings of a 5-point Likert scale question from our initial screening questionnaire (1 = not at all familiar, 5 = very familiar), the majority of participants (n = 33) claimed that they were moderately familiar with the concept of the metaverse (by choosing the choice number 4 on the Likert scale). Moreover, all participants claimed to be at least somewhat familiar with the concept of the metaverse. This can be surmised from the fact that no one opted for choice number 1 on the Likert scale. Furthermore, every participant was found to be familiar with at least one mainstream metaverse platform. Figure \ref{fig: dist1} shows the number of participants who were conversant with each of the mainstream metaverse platforms in 2022: i.e., \textit{Blocktopia} \cite{BloktopiaSite}, \textit{Cryptovoxels} \cite{CryptovoxelsSite}, \textit{Decentraland} \cite{DecentralandSite}, \textit{Meta Platforms} \cite{MetaPlatformsSite}, \textit{Mozilla Hubs} \cite{MozillaHubsSite}, \textit{Omniverse} \cite{NVIDIAOmniverseSite}, \textit{Roblox} \cite{RobloxSite}, \textit{Sandbox} \cite{SandboxSite}, \textit{Somnium Space} \cite{SomniumSpaceSite}, \textit{Spatial} \cite{SpatialSite}, and \textit{VRChat} \cite{VRChatsite}. 

\section{Supplementary Information on Interview Design and Procedures}\label{sup2}
Through an iterative process, our team collated 23 questions for the interviews. To this end, discussions were initially carried out internally with our research team, and subsequently, we had discussions with a few disabled students with whom we were acquainted. Next, we compared our potential list of questions to prior research studies belonging to the domains of human factors research, accessibility provisions, human-computer interaction, and L@S. We looked at papers in which interviews were the primary source of qualitative data (e.g., \cite{10.1145/3313831.3376581, 10.1145/3441852.3471231, 10.1145/3173574.3173799, 10.1145/3491102.3517574}). Finally, methodological resources were consulted (refer to \cite{blandford2016qualitative}). 

Table \ref{tab:interview_questions} illustrates the final set of questions set out across the four salient themes: \textit{warm-up questions}, \textit{a fictional classroom design}, \textit{personal concerns}, and \textit{design considerations}.  

\begin{table*}[t!]
  \centering
  \caption{The interview questions used in this study. All interviews were semi-structured, with follow-up questions asked as needed.}
  \small
    \begin{tabular}{cp{14cm}}
    \toprule
    $\star$ & \textbf{Warm-up Questions (to create a welcoming environment and help participants focus)} \\
    1     &  How do you see higher education's future in the metaverse? \\
    2     & Would [participant's name] be interested in attending a metaverse classroom? Could you please elaborate further?  \\
    3     & [Participant's name], have you used any metaverse platforms in the past? If yes, could you please tell me the names of the platforms and explain the reasons for choosing them?   \\
    4     & Please elucidate the advantages and disadvantages of the existing metaverse platforms in general and metaverse classrooms in particular. \\
    5     & How far are current immersive technologies from making the metaverse accessible and inclusive to all? In your opinion, what impact would this have on the educational opportunities for students with disabilities? \\
    \midrule
    $\star$ & \textbf{Designing a Fictional Metaverse Classroom} \\
    6     & Metaverse-based education is still in its infancy. Consider you are tasked with creating a brand new metaverse classroom that is accessible and inclusive to all, especially students with disabilities. What is going to be different in your classroom? \\
    7     & What will be the equipment, tools, and software that your classroom will comprise? Could you please describe them? \\
    8     & Can you please explicate some of your activities in this new classroom? \\
    9     & What do avatars and NFT objects look like in your classroom? How can people design, build, or customize them? \\
    10     & Is [participant's name] in metaverse you? What has changed? \\
    11    & How would you define the relationship between this virtual classroom (and avatars) with your actual physical environment (and physical body)? \\
    12    & Where are the teachers, caregivers, and other students in your classroom? What do they do? How do you interact with them and vice versa? \\
    13    & What are the new assistive technologies found in your metaverse classroom? How do you interact with your current assistive technologies? \\
    14    & How does your classroom improve the experience of students with disabilities? \\
    \midrule
    $\star$ & \textbf{Concerns} \\
    15    & What inclusion and accessibility barriers persist in your fictional classroom? Why do they still exist? \\
    16    & Do you have any new concerns about your novel classroom design? How can you possibly avoid or mitigate them? \\
    \midrule
    $\star$ & \textbf{Design Considerations} \\
    17    & What are the impediments to the creation of such a classroom? \\
    18    & How can we overcome those impediments? \\
    19    & Consider you want to assist a metaverse developer (or engineer) in creating your ideal metaverse classroom. What is the most important piece of advice you would give to this developer? \\
    20    & How can this developer create a more accessible and inclusive classroom like the one you described? \\
    21    & What could possibly go wrong with this new arrangement? \\
    22    & What can be done to avoid or minimize miscommunications? \\
    23    & Do you have any closing remarks? \\
    \bottomrule
    \end{tabular}%
  \label{tab:interview_questions}%
\end{table*}%

In line with the previous research conducted by Mogavi et al. \cite{mogavi2022promotional, hadi2022gamification}, participants were handed over a copy of the interview questions at least one day prior to their interviews to have sufficient time to get acquainted with the questions and (if they wanted or needed) choose the best way to present their responses (in a more comfortable manner). A free online meeting scheduling tool called \textit{Doodle} \cite{DoodleWebsite} helped determine the interview dates and times, taking cognizance of the interviewees' time preferences and the availability of all interviewers (the first three authors). 

Interviews were carried out in English (including ASL or using text\footnote{If requested by the participant}). On the day when the interviews were supposed to be held, the participant and their authorized caregivers, if present, were greeted cordially by one of the interviewers. It was reiterated that the participants could exercise the option to quit the study at any time without any questions, penalties, restrictions, or negative ramifications. we also spent a certain amount of time reviewing the format, structure, length, and purpose of our study together with our participants and their caregivers. The caregivers were allowed to accompany the participants during the entire process of the interview and if necessary, assist them (in any way that was needed). 

The third author took sufficient time to confirm that the participant had comprehended the consent form in its entirety. To do so, the third author went through a checklist similar to the one created by \cite{10.1145/3411764.3445150}. Each interview commenced only after obtaining a second confirmatory consent (live), accompanied by audio or video recording on the day of the interview. According to the second and third authors' professional experience, this reconfirmation on interview days provides an additional layer of protection, particularly for certain groups of disabled participants (who are more vulnerable to abuse or social pressure), and thus should not be considered a waste of time.  

The duration of each interview was between 50 to 75 minutes, with an average of 60 minutes. We followed the same protocol with each participant, going through all the questions in Table \ref{tab:interview_questions} in the order listed. Follow-up questions were also asked sporadically. In addition, some questions were revised if the information unraveled therefrom necessitated it. All interviews were recorded (textually and via audio and video), with all non-ASL interviews being automatically transcribed and rectified by the authors. The second and third authors manually transcribed all ASL-based interviews (n = 12). Upon the data collection’s conclusion, each participant was given an Amazon gift card worth \$30 to express our appreciation for their participation.

\section{Supplementary Information on Data Analysis}\label{sup3}
All in all, it took four rounds to complete the iterative process of coding. To begin with, each coder put in two weeks to carefully study the interview transcripts (reportedly more than once) to familiarize themselves with the data. Thereafter, each coder independently open-coded the data for a duration of one week and exchanged their initial set of codes with each other in preparation for the subsequent discussion meetings. After holding multiple 1-hour discussion meetings via Zoom, the coders settled in on nine initial codes.

The second and third iterations were very similar to the first one but involved more in-depth analysis and extended discussions to help get the initial codebook refined and condensed. In the second and third rounds of coding, the number of codes was reduced to seven and then six, respectively. Regarding the third and fourth iterations, Cohen's kappa measure was used to calculate inter-rater reliability to assess the agreement in the coding as well as to identify inconsistencies or ambiguities in the final codebook. In the third round, the inter-rater reliability was calculated at 0.54; however, this was too low to indicate satisfactory agreement (see \cite{lazar2017research}). For this reason, the coders met with the other authors and some interviewees (who were available) to figure out how to fix the problem of confounding codes. After two consequent 1-hour meetings with everyone via Zoom, the issues were resolved for the final codebook. Both coders finally agreed on five main (recurring) themes to be reported in this paper: i.e., Recognition, Empowerment, Engagement, Privacy, and Safety. The inter-rater reliability measure for the last coding round (4th iteration) was 1, indicating a complete agreement between both coders \cite{lazar2017research}.

\end{document}